%% file: arxiv.tex
\documentclass[3p]{elsarticle}
\usepackage{epstopdf}
\usepackage{subfigure,subcaption,graphicx,float}
\usepackage{amsmath, amsfonts, amssymb}
\usepackage{amsthm}
\usepackage{yhmath}
\usepackage{epsf}
\usepackage{amsfonts}
\usepackage{stmaryrd}
\usepackage{amssymb}
\usepackage{leftidx}
\usepackage{xcolor}
\usepackage{mathtools}
\usepackage{placeins}
\usepackage{booktabs}
\usepackage{enumitem}
\usepackage{caption}
\usepackage{multirow}
\usepackage{stackengine}
\usepackage[utf8]{inputenc}
\usepackage[english]{babel}
\usepackage{bm}
\usepackage{array}
\usepackage{siunitx}
\usepackage{url}
\usepackage{algorithm}
\usepackage{algpseudocode}

\let\caronaccent\v   
\input{definitions}
\let\v\caronaccent

\algrenewcommand\algorithmicrequire{\textbf{Input:}}
\algrenewcommand\algorithmicensure{\textbf{Output:}}

\usepackage{listings}
\usepackage{xcolor}

\newcommand{\R}{\mathbb{R}}
\newcommand{\hmin}{h_{\min}}
\newcommand{\Ibar}{\bar{\mathcal{I}}}
\newcommand{\Tcal}{\mathcal{T}}
\newcommand{\Acal}{\mathcal{A}}

\theoremstyle{plain}

\long\def\symbolfootnote[#1]#2{\begingroup%
\def\thefootnote{\fnsymbol{footnote}}\footnote[#1]{#2}\endgroup}

\begin{document}
\begin{frontmatter}

\title{Adaptive Crack-Tip Localization of Phase-Field Fracture Computations  \vspace{0.1cm}}

\vspace{-0.1cm}

\author[GT]{Aarosh Dahal}
\ead{adahal8@gatech.edu}

\author[GT]{Aditya Kumar\corref{cor1}}
\ead{aditya.kumar@ce.gatech.edu}

\address[GT]{School of Civil and Environmental Engineering, Georgia Institute of Technology, Atlanta, GA 30332, USA \vspace{0.05cm}}


\cortext[cor1]{Corresponding author}

\begin{abstract}
\vspace{-0.1cm}

Phase-field models of fracture provide a unified framework for simulating crack nucleation and propagation, but their computational cost remains high because the governing equilibrium and phase-field equations are typically iteratively solved over the entire domain at every staggered iteration. We present an extension of the recently developed adaptive mesh refinement and coarsening framework that localizes not only spatial resolution but also repeated solution effort to a small active patch around the crack tip for linear elastic brittle materials. The inactive exterior is eliminated through a Schur complement and condensed onto the patch interface, and its factorization is reused as long as the patch remains fixed. The phase field is frozen in the exterior, introducing only a small approximation in the crack wake. The method is implemented in FEniCSx and evaluated for quasi-static crack propagation, dynamic crack branching, three-dimensional mixed-mode crack growth, and coalescence of multiple cracks in a unit cell. Across these problems, the condensed formulation reproduces the reference solutions with differences in global quantities below 0.5\%, while substantially reducing the number of active degrees of freedom. In the quasi-static surfing problem, wall time speedups increase from 6.9 to 22.8 as the domain size is doubled, demonstrating increasing computational advantage for large-scale phase-field fracture simulations.

\endkeyword

\end{abstract}
\end{frontmatter}

\section{Introduction}\label{Sec:Introduction}

The phase-field framework for fracture provides a convenient approach for incorporating the complete physics of nucleation and propagation in brittle materials and solving the resulting equations using finite elements \cite{Francfort98, Bourdin00, KBFLP20}. The crack surface is represented by a continuous phase field that varies smoothly from intact ($v=1$) to fractured ($v=0$) material over a region whose width is governed by the regularization length $\varepsilon$. Accordingly, the finite element mesh must resolve $\varepsilon$ with sufficient spatial resolution, typically $h \sim \varepsilon/5$, in regions through which cracks propagate \cite{Tanne18}. Conventionally, this requirement has been assumed to apply throughout the entire region occupied by a phase-field crack.

Recent work \cite{dahal2026amrc}, however, has shown that the regularization length need only be fully resolved within a small neighborhood of the crack tip, where the strength-constrained competition between potential and fracture energies governs crack evolution \cite{lopez2025Whenandwhere, WK2025, lopez2026nucleation}. In the crack wake, where the phase-field crack is fully formed and the phase field has ceased to evolve, the phase-field crack need only retain the mechanical behavior of a sharp crack and not necessarily its optimum profile associated with the crack-density functional. The mesh can therefore be coarsened in this region, with the phase-field crack represented as a crack band. This observation was used to develop an efficient adaptive mesh refinement and coarsening (AMRC) scheme in which the finely resolved region moves adaptively with the crack tip. The method was demonstrated to be robust for multiple crack tips, crack branching, mixed-mode propagation, and crack nucleation. A key ingredient underlying this robustness is a physics-based indicator for spatial adaptivity derived from the material strength. Because violation of the strength criterion is a necessary condition for crack evolution, the indicator naturally identifies the crack-tip region and distinguishes it from the crack wake. This eliminates the need to explicitly track the crack tip, introduce special treatments for nucleation or branching, or employ a predictor-corrector strategy to reposition the fine-mesh patch based on the computed phase-field solution.

The computational implications are substantial. The computationally expensive finely resolved degrees of freedom are concentrated near the crack tip and therefore depend primarily on the intrinsic fracture length scales of the material. For a  single propagating crack, the number of finely resolved degrees of freedom remains approximately constant, independent of both crack length and overall domain size. The computational cost of large structural problems can therefore be reduced substantially compared with existing approaches that retain a fine mesh along the entire crack path and struggle to consistently identify nucleation and crack-tip regions due to use of heuristic indicators. These results motivate a further question: if the majority of the degrees of freedom can be localized near the crack tip, can the computational effort required to solve the governing equations be localized there as well?\footnote{The idea of concentrating computational resolution near a crack tip has a classical precedent in the boundary-layer formulation of fracture mechanics \cite{rice1968path, rice1968mathematical}.} The present work addresses this question.

The AMRC indicator naturally separates the crack-tip region from the crack-wake region. The crack wake contains fully formed phase-field crack bands in which the phase field undergoes little or no further evolution. Once the phase field in the fully developed crack wake drops near zero, irreversibility prevents healing, and its subsequent evolution becomes negligible. Consequently, for linear problems, the elastic stiffness associated with the crack-wake region remains essentially unchanged.
Outside the crack-tip and crack-wake regions, the phase field remains exactly 1 for the compact phase-field formulations such as the AT$_1$ formulation \cite{pham2011gradient, Tanne18}. Therefore, for linear elastic brittle materials, the repeated algebraic solution can be effectively reduced to the crack-tip region provided that the active crack-tip patch remains fixed. This observation forms the central premise of the present work.

For a fixed crack-tip patch, the displacement degrees of freedom in the exterior region can be statically condensed, through a Schur complement, onto the interface with the active patch \cite{przemieniecki1963matrix, wilson1974static, toselli2005domain}. For problems in which the crack evolves slowly over several hundred staggered iterations, and potentially over multiple load steps while remaining within the same active patch, the same factorization of the stiffness matrix associated with the exterior region can therefore be reused. Similarly, because the phase-field degrees of freedom in the exterior region remain nearly fixed, they can be kept frozen. If the phase field within the crack wake has not reached zero before the active patch advances, this condensation introduces a small approximation error. The results presented here demonstrate, however, that this error remains negligible for brittle fracture problems under both quasi-static and dynamic loading.

Explicit formation of the Schur complement introduces dense coupling among the interface degrees of freedom, with storage and algebraic costs that increase with the interface size. In the present formulation, however, this dense contribution is restricted to the boundary of the active patch, while the interior active-region operator retains its sparse finite element structure. Moreover, because the active patch translates with a propagating crack without increasing in size, the number of interface degrees of freedom remains approximately constant. Explicit condensation therefore remains computationally tractable even as the crack length and exterior domain size increase.

The algorithm for AMRC and condensation of the linear inactive parts is described in Section \ref{Sec:AMR}, following an introduction to the phase-field theory in Section \ref{Sec:Theory}. Results are presented in Section \ref{Sec:Results} for two- and three-dimensional crack growth under quasi-static loading, crack branching under dynamic loading (through Schur-complement condensation of the effective dynamic operator), and the growth and coalescence of multiple cracks in a unit cell. For each problem, we quantify the number of active nodes and assess solution accuracy against the corresponding reference solution obtained without condensation. Section \ref{Sec:Conclusions} provides concluding remarks.

\section{The phase-field fracture theory}\label{Sec:Theory}

A brief description of the phase-field approach of Kumar et al.\ \cite{KFLP18, KBFLP20} is discussed in this section. Other approaches have been proposed in the literature to model nucleation and propagation of fracture in linear elastic brittle materials \cite{Tanne18, wu2017} and adaptive mesh refinement algorithms without coarsening have been proposed based on different indicators \cite{Gupta2022_AdaptiveMeshRefinement}. The proposed algorithm in Section \ref{Sec:AMR} applies to all of them as long as a strength surface is well defined in that approach. 

\subsection{Phase-field model for brittle materials under quasi-static loading}

Consider an elastic body occupying a domain $\Omega \subset \R^N$ ($N = 2$ or $3$) in its reference configuration. Under external loads,  the body experiences a displacement field $\bfu(\bfX,t)$. We write the infinitesimal strain tensor as
\begin{equation*}
\bfE(\bfu)=\dfrac{1}{2}(\nabla\bfu+\nabla\bfu^T).
\end{equation*}
and the isotropic linear elastic stored-energy function as
\begin{equation}
	W(\bfE(\bfu)) =\mu \, {\rm tr}\,\bfE^2+\dfrac{\lambda}{2}({\rm tr}\,\bfE)^2,\label{W-mu}
\end{equation}
where $\mu$ and $\lambda$ are the Lam\'e constants. Recall the basic relations $\mu=E/(2(1+\nu))$ and $\lambda=E\nu/((1+\nu)(1-2\nu))$, where $E$ is the Young's modulus and $\nu$ is the Poisson's ratio. The stress-strain relation is given by
\begin{equation*}
	\boldsymbol{\sigma}(\bfX,t)=\dfrac{E}{1+\nu}\bfE+\dfrac{E\,\nu}{(1+\nu)(1-2\nu)}({\rm tr}\,\bfE)\bfI.
\end{equation*}
Subject to the appropriate initial and boundary conditions, and in the absence of body forces and inertia, the displacement field $\bfu_k(\bfX)=\bfu(\bfX,t_k)$ and the phase field $v_k(\bfX)=v(\bfX,t_k)$ at any material point $\bf X\in\overline{\mathrm{\Omega}}$ and discrete time $t_k\in\{0=t_0,t_1,...,t_m,$ $t_{m+1},...,$ $t_M=T\}$ are obtained from solving system of coupled partial differential equations (PDEs)
{
\begin{equation}
\left\{\begin{array}{lll}
\hspace{-0.15cm} {\rm div}\left[v_{k}^2 \dfrac{\partial W}{\partial \bfE}(\bfE(\bfu_{k}))\right]=0 ,
\vspace{0.1cm}\\
\hspace{-0.15cm}
\dfrac{3}{4} \varepsilon \, \de \,  G_c \triangle v_{k}=2 v_{k}\, W(\bfE(\bfu_{k}))  +c_\texttt{e} -\dfrac{3}{8} \dfrac{ {\delta^\varepsilon}  \, G_c }{\varepsilon}, & \text{if } 0<v_k<\overline{v}_k,
\vspace{0.1cm}
\end{array}\right. \label{phase-field-equations}
\end{equation}
where $\varepsilon$ is the regularization length, and $G_c$ is the fracture toughness. The second relation in (\ref{phase-field-equations}) states the evolution equation for the phase field as a bound-constrained problem. 
The phase field is restricted to the admissible range $0\le v_k\le\overline{v}_k$,
where the upper bound
\begin{equation}
\overline{v}_k(\bfX)=\left\{\begin{array}{ll} v_{k-1}(\bfX), & \text{if } v_{k-1}(\bfX)\le v_a,\\[4pt] 1, & \text{otherwise},\end{array}\right.
\label{eq:bounds}
\end{equation}
enforces the irreversibility of fracture: at material points where the phase field has dropped below the threshold $v_a=0.05$, it is not allowed to increase again, so that a fully developed crack cannot heal. Implementing weak irreversibility in this manner allows the crack to fully develop and yield the most physical results \citep{dolbow2025uniform}. The equations (\ref{phase-field-equations}) are consistent with a strength-constrained variational formulation recently defined for sharp cracks \cite{lopez2026nucleation}.

}

The constitutive prescription for the driving force $c_\texttt{e}(\bfX,t)$ in (\ref{phase-field-equations})$_2$ depends on the strength surface of the material. The formulation allows for an arbitrary choice of the strength surface as shown recently by Chockalingam et al. \cite{chockalingam2025MCHB}. In this work, we have adopted the Drucker-Prager (DP) surface, expressed as follows:
\begin{equation}
	\mathcal{F}(\bfS)=\sqrt{J_2}+\gamma_1 I_1+\gamma_0=0 ,\label{DP}
\end{equation}
with
\begin{equation}
	\gamma_0=-\sss, \qquad  \gamma_1=\dfrac{\sqrt{3} \sss-\sts}
{\sqrt{3}\sts},
\end{equation}
where
\begin{equation}
	I_1={\rm tr}\,\boldsymbol{\sigma}\qquad {\rm and}\qquad  J_2=\dfrac{1}{2}{\rm tr}\,\boldsymbol{\sigma}^2_{D}\qquad {\rm with}\quad \boldsymbol{\sigma}_{D}=\boldsymbol{\sigma}-\dfrac{1}{3}({\rm tr}\,\boldsymbol{\sigma})\bfI\label{T-invariants}
\end{equation}
stand for two of the standard invariants of the stress tensor $\boldsymbol{\sigma}$, while the constants $\sts>0$ and $\sss>0$ denote the uniaxial tensile and shear strengths of the material, respectively.
Corresponding to the DP surface, the functional form for $c_\texttt{e}(\bfX,t)$ was presented in \cite{KRLP22}:
\begin{align}
c_{\texttt{e}}(\bfX,t)=\beta_2^\varepsilon\sqrt{\mathcal{J}_2}+\beta_1^\varepsilon \mathcal{I}_1-
v (1-sgn(\mathcal{I}_1)) \, W(\bfE),\label{cehat-2022}
\end{align}
with
\begin{equation}
\left\{\begin{array}{l}
\beta^\varepsilon_1=\dfrac{1}{\shs}\delta^\varepsilon\dfrac{G_c}{8\varepsilon}-\dfrac{2\mathcal{W}_{\texttt{hs}}}{3\shs}\vspace{0.2cm}\\
\beta^\varepsilon_2=\dfrac{\sqrt{3}(3\shs-\sts)}{\shs\sts}\delta^\varepsilon\dfrac{G_c}{8\varepsilon}+
\dfrac{2\mathcal{W}_{\texttt{hs}}}{\sqrt{3}\shs}-\dfrac{2\sqrt{3}\mathcal{W}_{\texttt{ts}}}{\sts}\end{array}\right.  \label{betas}
\end{equation}
where 
$$\mathcal{W}_{\texttt{ts}}= \dfrac{\sts^2}{2 E}, \quad \mathcal{W}_{\texttt{hs}}= \dfrac{\shs^2}{2 \kappa}, \quad \shs=\dfrac{2 \sts  \scs} { 3 (\scs - \sts)}, \quad \kappa= \dfrac{E}{3 (1 - 2 \nu)}.$$
and 
$\mathcal{I}_1$ and $\mathcal{J}_2$ stand for the invariants (\ref{T-invariants}) of the degraded Cauchy stress
\begin{equation*}
\boldsymbol{\sigma}(\bfX,t)=v^2\dfrac{\partial W}{\partial \bfE}(\bfE(\bfu))
\end{equation*}
The coefficient $\delta^{\varepsilon}$ is given by
\begin{equation}
\delta^{\varepsilon} = \left(1 + \frac{3h}{8\varepsilon}\right)^{-2} \left(\frac{\sts + (1 + 2\sqrt{3})\shs}{(8 + 3\sqrt{3})\shs}\right) \frac{3G_c}{16\mathcal{W}_{\texttt{ts}}\varepsilon} + \left(1 + \frac{3h}{8\varepsilon}\right)^{-1} \frac{2}{5},
\label{eq:delta}
\end{equation}
where $h$ denotes the finite element size.

\subsubsection{The phase-field strength surface indicator}\label{Sec:Indicator}

The governing equations (\ref{phase-field-equations}) dictate \cite{KBFLP20} that the phase field can only evolve from its initial value of $v=1$ when,
\begin{equation}
\mathcal{I}(\mathbf{X}, t) = 2 \, W(\bfF) + \ce - \dfrac{3 \de G_c}{8 \eps}=0 \label{strength-surface-ce}
\end{equation}
obtained by setting $v=1$ in the right-hand side of the evolution equation for phase-field (\ref{phase-field-equations})$_2$. {The surface $\mathcal{I}(\mathbf{X}, t)=0$} can be regarded as a phase-field approximation of the strength surface ${\mathcal{F}}(\boldsymbol{\sigma})$ (\ref{DP}) and it signals the onset of fracture nucleation.
The phase-field strength surface reduces to the exact strength surface for $\eps \searrow 0$ \cite{KBFLP20}.
Dahal et al. \cite{dahal2026amrc} utilize this fact to base the indicator for adaptive mesh refinement and coarsening on $\mathcal{I}(\mathbf{X}, t)$, as discussed further in Section \ref{Sec:AMRC}.

\subsubsection{The finite element formulation}\label{Sec:FEM-QS}

The finite element formulation of the governing equations in the phase-field approach to fracture is now well established in the literature; see, for instance, Section 4 in \cite{KFLP18} for the implementation of equations \eqref{phase-field-equations}. We briefly recall the formulation for only the equilibrium equation for later use in Section \ref{Sec:Condensation}.

We define the trial function space $\mathcal{S}$ and  test function spaces $\mathcal{W}$ for the displacement field as 
\begin{equation}
\mathcal{S} = \{ \mathbf{u} \in (H^{1}(\Omega))^{d} \mid \mathbf{u} = \bar{\mathbf{u}} \text{ on } \partial\mathrm{\Omega}_\mathcal{D} \},
\notag
\end{equation}
\begin{equation}
\mathcal{W} = \{ \mathbf{w} \in (H^{1}(\Omega))^{d} \mid \mathbf{w} = \mathbf{0} \text{ on } \partial\mathrm{\Omega}_\mathcal{D} \},
\notag
\end{equation}
The weak form of the force balance \eqref{phase-field-equations}$_1$ is given by :
\begin{equation}
\left\{\begin{array}{ll}
\text{Find } \mathbf{u} \in \mathcal{S}, \\[8pt]
 \int_{\Omega} v^2 \boldsymbol{\sigma}(\mathbf{u}) : \mathbf{E}(\mathbf{w})\,dx 
=  \int_{\partial \Omega} \mathbf{\bar{t}} \cdot \mathbf{w}\,ds,\quad \forall\, \mathbf{w}\in \mathcal{W}, \\[8pt]
\end{array}\right. 
\label{weakBVP-u-theory}
\end{equation}
where $\mathbf{\bar{t}}$ is the applied traction. Next, we consider partitions of the domain $\Omega$ into non-overlapping simplicial elements. We define conforming finite element spaces $\mathcal{S}^{h} \subset \mathcal{S}$ and $\mathcal{W}^{h} \subset \mathcal{W}$. The Galerkin form of the problem is stated as follows: find $\mathbf{u}^h \in \mathcal{S}^{h}$ such that for all $\mathbf{w}^h \in \mathcal{W}^{h}$:
\begin{equation} 
 \int_{\Omega} (v^h)^2 \boldsymbol{\sigma}(\mathbf{u}^h) : \mathbf{E}(\mathbf{w}^h) dx =  \int_{\partial\Omega} \bar{\mathbf{t}} \cdot \mathbf{w}^h  ds ,
\end{equation}
The approximation for the displacement field $\mathbf{u}^h$, phase field $v^h$ and the test function $\mathbf{w}^h$ in the Galerkin method are given by:
\begin{equation} 
\mathbf{u}^h(\mathbf{X}, t) = \sum_{i=1}^{N_e} N(\mathbf{X}) \tilde{\mathbf{u}}(t), \quad v^h(\mathbf{X}, t) = \sum_{i=1}^{N_e} N(\mathbf{X}) \tilde{v}(t), \quad \mathbf{w}^h(\mathbf{x}) = \sum_{i=1}^{N_e} N(\mathbf{X}) \tilde{\mathbf{w}},
\label{uh-wh}
\end{equation}
where $N(\mathbf{X})$ are linear shape functions, $N_e$ represents number of nodes, $\tilde{\mathbf{u}}$ are the nodal displacements and $\tilde{\mathbf{w}}$ are the corresponding variations. The resultant linear discrete form of the equilibrium equation becomes 
\begin{equation}
    \mathbf{K} \tilde{\mathbf{u}} = \mathbf{f}, 
\end{equation}
where the components of the stiffness matrix $\mathbf{K}$, and external force vector $\mathbf{f}$ are:
\begin{align}
    \mathbf{K} &= \int_{\Omega} (v^h)^2 \mathbf{B}^T \mathbf{D} \mathbf{B} \, dx, \\ 
    \mathbf{f} &= \int_{\partial\Omega} N \bar{\mathbf{t}} \, ds
    \label{K-f}
\end{align}
Here, $\mathbf{D}$ is the elasticity matrix in Voigt representation and $\mathbf{B}$ is the strain-displacement matrix. Note that the stiffness matrix explicitly depends on the phase-field $v^h$.

\subsection{Phase-field model for dynamic fracture in brittle materials}

A simple extension of the quasi-static brittle fracture model \eqref{phase-field-equations} is to add the inertial term to the balance of linear momentum and keeping everything else unchanged \cite{liu2024effects}. Such a formulation agrees well with a host of benchmark problems such as the Kalthoff-Winkler test and the dynamic branching test. The modified formulation stated in linear elasticity for which it has previously been evaluated is as follows:
subject to the appropriate initial and boundary conditions, the displacement field $\mathbf{u}_k(\mathbf{X})=\mathbf{u}(\mathbf{X},t_k)$ and phase-field $v_k(\mathbf{X})=v(\mathbf{X},t_k)$ at any material point $\mathbf{X}\in\overline{\Omega}$ and discrete time $t_k\in\{0=t_0,t_1,\dots,t_m,$ $t_{m+1},\dots,$ $t_M=T\}$ are determined by the system of coupled partial differential equations
\begin{equation}
\left\{\begin{array}{ll}
{\rm div}\left[v_{k}^2 \dfrac{\partial W}{\partial \bfE}(\bfE(\bfu_{k}))\right]
= v_{k}^2 \,\rho\,\ddot{\mathbf{u}}_{k},\\[10pt]
\dfrac{3}{4} \varepsilon \, \de \,  G_c \triangle v_{k}=2 v_{k} W(\bfE(\bfu_{k}))+c_\texttt{e}(\bfX,t_{k})- \dfrac{3}{8}  \dfrac{\de \, G_c}{\varepsilon}, \\[10pt]
\end{array}\right.
\label{phase-field-dynamics}
\end{equation}
{where the phase-field equation is  subject to the bounds (\ref{eq:bounds}), and} $\ddot{\mathbf{u}} = {\partial^2 \mathbf{u}}/{\partial t^2}$ stands for the acceleration field. In this formulation, the mass density is degraded so that the fractured region does not get artificially distorted and widen \cite{durussel2026dynamic, del2026dynamic, dahal2026dynamic}.
Since in the approach of Kumar et al., the regularization length $\varepsilon$ is a free parameter, Dahal et al. \cite{dahal2026dynamic} recently argued that degrading density only introduces a regularization error in the conservation of mass on par with the error introduced in elastic behavior by smearing a sharp crack over a finite length scale. The phase field strength surface indicator remains the same as in the quasi-static formulation.

\subsubsection{The finite element formulation}\label{Sec:FEM-Dyn}

The complete finite element formulation of the dynamic fracture governing equations \eqref{phase-field-dynamics} has been presented recently in \cite{dahal2026dynamic}. Again, for the sake of later use in Section \ref{Sec:Condensation}, we briefly recall the FE formulation for the equilibrium equation.

The weak form of the force balance \eqref{phase-field-dynamics}$_1$ is:
\begin{equation}
\left\{\begin{array}{ll}
\text{Find } \mathbf{u} \in \mathcal{S}, \\[8pt]
 \displaystyle \int_{\Omega} v^2 \rho\, \ddot{\mathbf{u}} \cdot \mathbf{w}\,dx 
+\int_{\Omega} v^2 \boldsymbol{\sigma}(\mathbf{u}) : \mathbf{E}(\mathbf{w})\,dx 
=  \int_{\partial \Omega} \mathbf{\bar{t}} \cdot \mathbf{w}\,ds,\quad \forall\, \mathbf{w}\in \mathcal{W}, \\[8pt]
\end{array}\right. 
\label{weakBVP-u-theory-dyn}
\end{equation}
and the Galerkin form is: 
\begin{equation} 
\int_{\Omega} (v^h)^2 \rho \ddot{\mathbf{u}}^h \cdot \mathbf{w}^h  dx + \int_{\Omega} (v^h)^2 \boldsymbol{\sigma}(\mathbf{u}^h) : \mathbf{E}(\mathbf{w}^h) dx =  \int_{\partial\Omega_{\mathcal{N}}} \bar{\mathbf{t}} \cdot \mathbf{w}^h  ds ,
\end{equation}
where  $\bfu^h$, $v^h$ and $w^h$ are according to \ref{uh-wh}. The semi-discrete form of the equilibrium equation becomes 
\begin{equation}
    \mathbf{M} \ddot{\tilde{\mathbf{u}}} + \mathbf{K} \tilde{\mathbf{u}} = \mathbf{f}(\tilde{v}),
\end{equation}
where the mass matrix $\mathbf{M}$ is
$$\mathbf{M} = \int_{\Omega} (v^h)^2 \rho N N^T \, dx,$$
and the stiffness matrix and force vector are given by (\ref{K-f}). We use HHT-$\alpha$ method \cite{Hilber1977_ImprovedDissipation} for temporal discretization. We discretize the time interval $[0,T]$ into $N{+}1$ steps 
$0=t_0 < t_1 < \cdots < t_N < t_{N+1}=T$  
with uniform time step $\Delta t = T/(N+1)$. In this method, the spatially discretized dynamic equilibrium equation is evaluated at the intermediate time $t^{n+1-\alpha_f}$:
\begin{equation}
  \mathbf{M}\,\ddot{\tilde{\mathbf{u}}}_{\,n+1}
  + \mathbf{K}\,\tilde{\mathbf{u}}_{\,n+1-\alpha_f}
  =
  \mathbf{f}(t_{\,n+1-\alpha_f}),
  \label{eq:time-discretize-equilibrium}
\end{equation}
where
\begin{equation}
  \tilde{\mathbf{u}}_{n+1-\alpha_f}
  =
  (1-\alpha_f)\,\tilde{\mathbf{u}}_{n+1}
  + \alpha_f\,\tilde{\mathbf{u}}_n.
\end{equation}
Using Newmark's update relations and simplifying, equation (\ref{eq:time-discretize-equilibrium}) can be reduced to the following linear system
\begin{equation}
  \mathbf{\bar{K}}\,\tilde{\mathbf{u}}_{n+1}
  =
  \mathbf{\bar{f}}
  ,
\end{equation}
with the effective stiffness matrix
\begin{equation}
  \mathbf{\bar{K}}
  =
  (1-\alpha_f)\,\mathbf{K}
  + m_1\,\mathbf{M}.
  \label{Eff-stiff-Dyn}
\end{equation}
and effective force vector
\begin{equation}
  \mathbf{\bar{f}}
  =
  \mathbf{f}(t_{\,n+1-\alpha_f})
  - \alpha_f\,\mathbf{K}\,\mathbf{u}_{n}
  + \mathbf{M}\!\left(m_1\mathbf{u}_{n} + m_2 \dot{\mathbf{u}}_{n} - m_3 \ddot{\mathbf{u}}_{n}\right).
  \label{Eff-force-Dyn}
\end{equation}
Here, $m_1={1}/({\beta\Delta t^2})$, $m_2={1}/({\beta\Delta t})$, and $m_3=({1-2\beta})/({2\beta})$ and $\beta$ is the Newmark constant chosen as $\beta=0.25(1+\alpha_f)^2$.

\section{Adaptive mesh refinement and coarsening with condensation of linear inactive parts}\label{Sec:AMR}

We first recall in Section \ref{Sec:AMRC} the adaptive mesh refinement and coarsening (AMRC) framework of \cite{dahal2026amrc}. We then present the algorithm for condensation of linear inactive parts (CLIP). In CLIP, the coupled problem is solved in a staggered manner only on a small patch of active cells around the crack tip, and the inactive remainder of the domain where stiffness remains constant is condensed onto the boundary of the patch (Sections \ref{Sec:Patch}--\ref{Sec:Move}). We call AMRC-CLIP the combination of AMRC with CLIP, in which the active patch follows the refined region around the crack tip. We call AMRC-CLIP-TW (tip and wake) a variant in which the patch also encloses the wake of the crack (Section \ref{Sec:TW}).

\subsection{Adaptive mesh refinement and coarsening (AMRC)}\label{Sec:AMRC}

\begin{figure}[h]
	\centering
	\includegraphics[width=6in]{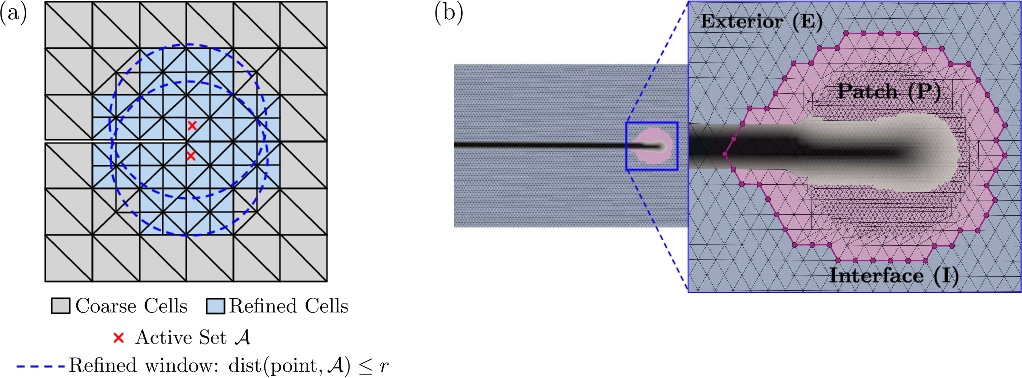}
	\caption{(a) AMRC introduced in \cite{dahal2026amrc}: the base mesh is refined within the distance $r$ of the active set $\Acal$, formed by the centroids of the active cells according to the criterion \eqref{eq:active}. (b) AMRC-CLIP introduced in this work: the patch $\mathcal{P}$ (pink) of phase field evolution defined based on the refined window, the exterior $\mathcal{E}$ with frozen phase field, and the interface (pink dots) onto which the exterior is condensed.}\label{Fig1}
\end{figure}

The AMRC algorithm presented in \cite{dahal2026amrc} operates on a fixed coarse \emph{base mesh} $\Tcal_0$ for ease of implementation in different finite element libraries where an option for de-refinement may not be available. Every mesh is obtained from the base mesh by $n$ levels of edge bisection down to $\hmin=h_0/2^n$ \cite{plaza2000mesh}. $\hmin$ is typically taken to be $\varepsilon/4$. The region of refinement is determined by the phase-field strength surface of Section \ref{Sec:Indicator} through the normalized indicator
\begin{equation}
\Ibar_{\texttt{c}}(\bfX,t)=\mathcal{H}(v-v_{\texttt{c}})\,\dfrac{2\,W+\ce\big|_{v=1}}{\dfrac{3}{8}\dfrac{\de G_c}{\varepsilon}}-1,
\label{eq:indicator}
\end{equation}
where $\mathcal{H}$ is the Heaviside function and $v_{\texttt{c}}=0.6$. A cell $K$ in the current mesh $\Tcal$  is considered \emph{active} when
\begin{equation}
\Ibar_{\texttt{c}}(\bfX_K,t)>-\theta,
\label{eq:active}
\end{equation}
where $\bfX_K$ is its centroid and $\theta=0.1$, that is, when the driving force reaches 90\% of the resistance. The set of the centroids of the active cells is defined as $\Acal$ as shown schematically in Fig.~\ref{Fig1}(a). The majority of phase field evolution takes place in and around these active cells. Therefore, the base mesh is refined to $\hmin$ within the distance $r=3\varepsilon$ of $\Acal$, shown in blue in Fig.~\ref{Fig1}(a) and referred to as the \emph{refined window}. A rectangular or oval-shaped window around the active cells can also be considered. 
Dahal et al. \cite{dahal2026amrc} showed that this method robustly identifies the region where a refined mesh is needed across a wide variety of boundary value problems, including problems where a new crack nucleates.

The robustness is derived from the fact that $\Ibar$ depends on the strength surface \eqref{strength-surface-ce} whose violation is a necessary condition for phase field evolution. The Heaviside factor further separates the crack-tip region from the crack-wake region, so the refined window remains restricted to a small volume. The insight presented in \cite{dahal2026amrc} was that the flanks of a fully developed phase-field crack are already unloaded; therefore, the Heaviside factor excludes its core, $v<v_{\texttt{c}}$, as well where the undegraded stress would otherwise violate the strength surface. 
The active cells thus form a thin crescent ahead of the crack tip. Importantly, with this indicator, the coordinates of the tip do not need to be explicitly tracked, which can become challenging in problems involving branching or multiple crack nucleation and growth.

Since every mesh is rebuilt from the base mesh around the current active set only, the crack wake returns to the base mesh (or alternatively a different mesh). The phase field crack in the wake is then represented by a band of fully degraded coarse elements, called the \emph{crack band}, which was shown in \cite{dahal2026amrc} to reproduce the global mechanical response of a fully resolved phase-field crack for $h_0\le2\varepsilon$. The crack-band representation was the second insight of \cite{dahal2026amrc}, recognizing that once the phase field stops evolving, a fine mesh resolving the regularization length scale is not needed.

AMRC thus keeps the number of elements nearly constant as a single crack grows. At each step, the coupled problem is solved by a staggered scheme that alternates between the displacement problem with $v$ fixed and the phase-field problem with $\bfu$ fixed. The mesh is rebuilt within this loop whenever an active cell is still coarse, and the refined window is then moved. In problems involving stable crack growth, the refined window may remain stationary for hundreds of staggered iterations as the phase field evolves slowly inside it.
Every staggered iteration still solves both problems on the entire domain, although the phase field evolves only in the refined regions. For linear elastic brittle materials, the stiffness in the region outside the refined window does not change. Therefore, it can be condensed onto the boundary of the patch once and reused in every staggered iteration. This is the idea of AMRC-CLIP.

\subsection{Splitting the domain into patch and exterior}\label{Sec:Patch}

We split the cells of the base mesh into a \emph{patch} $\mathcal{P}\subset\Tcal_0$ and an \emph{exterior} $\mathcal{E}=\Tcal_0\setminus\mathcal{P}$ (Fig.~\ref{Fig1}(b)). The nodes shared by the two form the \emph{interface} $\mathcal{I}$. The patch is a small set of base cells around the active cells enclosing the current refined window in which the phase field evolves. The cells of the previous refined window are also kept in the patch since the phase field typically continues evolving in an $\mathcal{O}(l_{ch})$ region behind the crack tip, where $l_{ch}$ is the characteristic fracture length scale in tension. Precisely, we say that the patch is a union of these two sets: 
\begin{equation}
\mathcal{P}=\mathcal{B}(\Acal, {R_{\texttt{p}}})\,\cup\, \mathcal{N}(\mathcal{D}\cap\mathcal{B}(\Acal_{\rm prev}, {R_{\texttt{p}}})),
\label{eq:patch}
\end{equation}
where
\begin{equation}
\mathcal{B}(\Acal, R)=\left\{K\in\Tcal_0:\ \operatorname{dist}(\bfX_K,\Acal)\le R\right\},
\end{equation}
Here, $\mathcal{A}$ is the current active set, while $\mathcal{A}_{\rm prev}$ denotes the active set corresponding to the most recent preceding refined window construction. $\mathcal{D}$ denotes the base cells that have a node at which phase field $v<1-10^{-5}$, and $\mathcal{N}(\mathcal{S})$ denotes a set $\mathcal{S}$ of base cells together with all the base cells that share a node with it. Once constructed, the patch is reconstructed only when one of the conditions described in Section \ref{Sec:Move} is satisfied.

Different choices can be made for the radius of the patch around the active cells $R_{\texttt{p}}$. A simple choice would be $R_{\texttt{p}}=r=3 \varepsilon$, where, recall, $r$ is the radius of the refined window. Another choice is to assume a larger $R_{\texttt{p}}$ so the crack can grow more before the patch needs to be reconstructed. This also means that the patch can cover a larger portion of the wake where the phase field may still be slightly evolving; in that case, the second set from \eqref{eq:patch} can also be potentially dropped. We make an intermediate choice of $R_{\texttt{p}}=4\varepsilon$ in this work, which shows good results. A typical active patch obtained in the simulations is shown in Fig.~\ref{Fig1}(b). The method can be altered to create a more regular-shaped patch if desired.

\subsection{Condensation of the exterior}\label{Sec:Condensation}

As shown in Section \ref{Sec:FEM-QS} and \ref{Sec:FEM-Dyn},  the discretized equilibrium problem is a linear system of the form $\mathbf{K}(v)\,\tilde{\mathbf{u}}=\mathbf{f}$ for the nodal displacements $\tilde{\mathbf{u}}$ with phase-field-dependent stiffness. We group the unknown nodal displacements into those of the patch nodes that are not on the interface ($P$), of the interface nodes ($I$), and of the remaining exterior nodes ($E$). The linear system reads in matrix form as
\begin{equation}
\begin{bmatrix}
\mathbf{K}_{PP} & \mathbf{K}_{PI} & \mathbf{0}\\[2pt]
\mathbf{K}_{IP} & \mathbf{K}^{\mathcal{P}}_{II}+\mathbf{K}^{\mathcal{E}}_{II} & \mathbf{K}_{IE}\\[2pt]
\mathbf{0} & \mathbf{K}_{EI} & \mathbf{K}_{EE}
\end{bmatrix}
\begin{Bmatrix}\tilde{\mathbf{u}}_P\\ \tilde{\mathbf{u}}_I\\ \tilde{\mathbf{u}}_E\end{Bmatrix}
=
\begin{Bmatrix}\mathbf{f}_P\\ \mathbf{f}_I\\ \mathbf{f}_E\end{Bmatrix},
\label{eq:block}
\end{equation}
where the superscripts $\mathcal{P}$ and $\mathcal{E}$ denote the contributions of the patch cells and of the exterior cells to the interface block. The right-hand side contains the external forces and the contributions of the prescribed displacements. The exterior blocks $\mathbf{K}_{EE}$, $\mathbf{K}_{EI}=\mathbf{K}_{IE}^T$, and $\mathbf{K}^{\mathcal{E}}_{II}$ depend only on the mesh and the phase field of the exterior. Both are frozen, so these blocks do not change until the patch moves. Eliminating
\begin{equation}
\tilde{\mathbf{u}}_E=\mathbf{K}_{EE}^{-1}\left(\mathbf{f}_E-\mathbf{K}_{EI}\,\tilde{\mathbf{u}}_I\right)
\label{eq:recover}
\end{equation}
yields the condensed system on the patch
\begin{equation}
\begin{bmatrix}
\mathbf{K}_{PP} & \mathbf{K}_{PI}\\[2pt]
\mathbf{K}_{IP} & \mathbf{K}^{\mathcal{P}}_{II}+\widehat{\mathbf{K}}_{II}
\end{bmatrix}
\begin{Bmatrix}\tilde{\mathbf{u}}_P\\ \tilde{\mathbf{u}}_I\end{Bmatrix}
=
\begin{Bmatrix}\mathbf{f}_P\\ \widehat{\mathbf{f}}_I\end{Bmatrix},
\qquad
\widehat{\mathbf{K}}_{II}=\mathbf{K}^{\mathcal{E}}_{II}-\mathbf{K}_{IE}\,\mathbf{K}_{EE}^{-1}\,\mathbf{K}_{EI},\quad
\widehat{\mathbf{f}}_I=\mathbf{f}_I-\mathbf{K}_{IE}\,\mathbf{K}_{EE}^{-1}\,\mathbf{f}_E.
\label{eq:condensed}
\end{equation}
This is the classical static condensation of structural analysis \cite{przemieniecki1963matrix, wilson1974static, toselli2005domain}. The Schur complement $\widehat{\mathbf{K}}_{II}$, the discrete Steklov--Poincar\'e operator of the exterior \cite{toselli2005domain}, is the exact stiffness with which the exterior resists a displacement of the interface. 
Equivalently, $\widehat{\mathbf{K}}_{II}$ is a discrete Dirichlet-to-Neumann (DtN) map of the exterior. DtN formulations have been used in fracture mechanics to couple a local nonlinear crack-tip region to a surrounding linear-elastic domain \cite{landis2011dtn}. Here, the DtN map is obtained directly from the finite element solution of the exterior domain.

When the displacements $\bar{\mathbf{d}}_B$ are prescribed on part of the exterior nodes, $\widehat{\mathbf{f}}_I=\mathbf{T}\,\bar{\mathbf{d}}_B$ with $\mathbf{T}=\mathbf{K}_{IE}\mathbf{K}_{EE}^{-1}\mathbf{K}_{EB}-\mathbf{K}^{\mathcal{E}}_{IB}$. 
Computing $\widehat{\mathbf{K}}_{II}$ and $\mathbf{T}$ requires one sparse Cholesky factorization of $\mathbf{K}_{EE}$ and one forward and backward substitution per interface degree of freedom. As long as the patch remains stationary, the same factorization can be used.

The condensed system holds for both quasi-static and dynamic problems; for dynamics, the effective stiffness matrix and force vector are defined by equations (\ref{Eff-stiff-Dyn})-(\ref{Eff-force-Dyn}). However, for dynamic problems, the condensed right-hand side $\widehat{\mathbf{f}}_I$ needs to be formed once per time step irrespective of whether the patch has moved since it depends on the displacement, velocity, and acceleration of the previous time step, which gets updated.

\subsection{Phase field in the exterior}\label{Sec:FrozenWake}

The phase field is considered frozen in the exterior and at the interface. Therefore, the phase-field problem \eqref{phase-field-equations}$_2$ or \eqref{phase-field-dynamics}$_2$ only need to be solved on the patch.
The freezing is exact away from the crack where the phase field stays at $v=1$ until the strength surface is reached for the chosen AT$_1$ regularization of surface energy.
In the crack band region in the wake, freezing is an approximation as the phase field may not have reached its minimum value of $v=0$. Accounting for the previously refined window in (\ref{eq:patch}) significantly reduces this effect. As shown in Section \ref{Sec:Results}, the resulting differences in the global responses remain below 0.5\%.
The error may be more significant for quasi-brittle phase field fracture models \cite{wu2017, bourdin2025variational} where the phase field evolves more slowly in the wake than in the brittle models. A larger refined window and/or patch size can be chosen for those models.

\subsection{Moving the patch}\label{Sec:Move}

The complete algorithm to move the patch is summarized in Algorithms \ref{alg:clip} and \ref{alg:move}.
The patch is updated in three situations:
\begin{enumerate}[leftmargin=*]
\item An active cell in the patch is coarse and the new refined window does not fit inside the current patch. This is checked in every staggered iteration after the solution of the equilibrium equation. 
\item A cell of the first layer of exterior cells becomes active. This is checked in every staggered iteration after the solution of the equilibrium equation and requires only the fields in one layer of elements instead of the complete exterior.
\item An exterior cell farther from the patch is active at the end of a step. This is checked when the exterior displacement is recovered from (\ref{eq:recover}) at the end of the step. The step is then not stopped. The patch is moved to include the cell, and the staggered iterations continue on the new patch. This is relevant for when a new crack may nucleate or under dynamic loading.
\end{enumerate}
%


\begin{algorithm}[h]
\caption{One loading step of AMRC-CLIP.}\label{alg:clip}
\begin{algorithmic}[1]
\Require patch $\mathcal{P}$ with its refined mesh and condensed exterior; fields $(\bfu_{k-1},v_{k-1})$
\Ensure fields $(\bfu_k,v_k)$
\State Apply the loading of $t_k$, form $\widehat{\mathbf{f}}_I$, and set the bounds (\ref{eq:bounds})
\Repeat
  \State Solve (\ref{eq:condensed}) for the displacement on the patch
  \State Find the active cells (\ref{eq:active}) in the patch and in the first exterior layer
  \If{an active cell is coarse} 
     \textsc{Move}
  \EndIf
  \State Solve the phase-field problem on the patch
\Until{the staggered iterations converge}
\State Solve (\ref{eq:condensed}) with the final phase field and recover $\tilde{\mathbf{u}}_E$ from (\ref{eq:recover})
\If{an exterior cell is active}
  \State \textsc{Move} and go to line 2
\EndIf
\end{algorithmic}
\end{algorithm}

\begin{algorithm}[h]
\caption{MOVE: Relocation of the patch around the active set $\Acal$.}\label{alg:move}
\begin{algorithmic}[1]
\State Recover $\tilde{\mathbf{u}}_E$ and copy the patch values at the base-mesh nodes to the fields on $\Tcal_0$
\State Select the new patch from (\ref{eq:patch}) and refine it around $\Acal$ as in AMRC
\State Factorize $\mathbf{K}_{EE}$; compute $\widehat{\mathbf{K}}_{II}$, $\mathbf{T}$, $\widehat{\mathbf{f}}_I$, and the rows of $\mathbf{K}_{EE}^{-1}\mathbf{K}_{EI}$ for the first exterior layer
\State Transfer the fields to the new patch
\end{algorithmic}
\end{algorithm}

\subsection{The tip-and-wake variant: AMRC-CLIP-TW}\label{Sec:TW}

The approximation of AMRC-CLIP can be removed by keeping the whole fractured region inside the patch. The exterior contains only intact material where $v=1$.  We call this AMRC-CLIP-TW. The refined window is still restricted to a small region around the crack tip. In AMRC-CLIP-TW, the patch is
\begin{equation}
\mathcal{P}_{\texttt{TW}}=\mathcal{B}(\Acal, {R_{\texttt{p}}})\,\cup\,\mathcal{N}(\mathcal{D}).
\label{eq:patch-tw}
\end{equation}
The price to pay is that the size of the patch grows with the length of the crack. We will evaluate this variant along with AMRC-CLIP in next section.

\subsection{Implementation in FEniCSx}\label{Sec:CLIPImplementation}

The method is implemented in FEniCSx \cite{barrata2023dolfinx, logg2012automated}, specifically DOLFINx version 0.9. It reuses the refinement and coarsening routines of AMRC discussed in detail in \cite{dahal2026amrc}. The matrix $\mathbf{K}_{EE}$ is factorized with the sparse Cholesky solver CHOLMOD \cite{chen2008cholmod}.
On the patch, the displacement problem is solved with the conjugate gradient method. It is preconditioned by the Cholesky factorization of the patch matrix of an earlier iteration, which is a good preconditioner because the phase field usually changes little between iterations. A direct solver can also be used. The phase-field problem is solved with the reduced-space active-set Newton method of PETSc \cite{balay2019petsc, benson2006flexible}.

\section{Numerical results}\label{Sec:Results}

We now assess AMRC-CLIP and AMRC-CLIP-TW algorithms on four problems of varying complexity:
\begin{itemize}[leftmargin=*]
\item steady quasi-static crack propagation in the surfing problem, where we also examine how the cost scales with the size of the domain (Section \ref{Sec:Surfing});
\item dynamic crack propagation, in which crack branches and the patch must follow two crack tips (Section \ref{Sec:Branching});
\item three-dimensional crack growth under in-plane shear, in which the crack kinks out of its initial plane, and the patch must enclose a 3D crack front (Section \ref{Sec:Shear3D});
\item the growth and coalescence of 16 cracks in a unit cell, in which many crack tip patches are active at once and merge with each other (Section \ref{Sec:Percolation}).
\end{itemize}
For each problem, the AMRC result without condensation and freezing of the phase field in the exterior is taken as a reference. It was shown in \cite{dahal2026amrc} that the AMRC result shows excellent agreement for the first two problems with the results obtained from uniformly refined meshes.
AMRC-CLIP-TW results should be in principle exact with respect to the reference result since they include the entire fractured region, which is the zone of nonlinearity, in their active patches.
AMRC-CLIP may introduce small errors, as we discussed before, because it may prematurely freeze the phase-field value in the crack wake. We will report this error for each problem with respect to the reference result.


To quantify the size of the problem, we report the number of \emph{active nodes}. These are all the nodes in the finite element mesh for AMRC and the nodes contained in the patch for AMRC-CLIP and AMRC-CLIP-TW. All three approaches share the same implementation and solvers. The material properties are listed in Table \ref{tab:params}. Unless stated otherwise, $h_0=\varepsilon$, $\hmin=\varepsilon/4$, $r=3\varepsilon$, $R_{\texttt{p}}=4\varepsilon$, $\theta=0.1$, $v_{\texttt{c}}=0.6$, and $v_a=0.05$. All computations were performed on a single core of an Intel Xeon Gold 6226 processor (2.7 GHz), with 16 GB of memory allocated to the job.

\begin{table}[h]
\centering
\caption{Material properties and regularization length used in the problems. Moduli and strengths are in MPa, $G_c$ in N/mm, and lengths in mm. $\sts$ is the uniaxial tensile strength, $\scs$ is the uniaxial compressive strength of the Drucker--Prager surface(\ref{DP}).}
\label{tab:params}
\small
\begin{tabular}{lcccccc}
\toprule
Problem & $E$ & $\nu$ & $G_c$ & $\sts$ & $\scs$ & $\varepsilon$ \\
\midrule
Surfing (\ref{Sec:Surfing}) & 9800 & 0.13 & 0.0911 & 27 & 77 & 0.175 \\
Dynamic branching (\ref{Sec:Branching}) & 72000 & 0.22 & 0.0038 & 6.16 & 18.5 & 0.625 \\
3D in-plane shear (\ref{Sec:Shear3D}) & 9800 & 0.3 & 0.0911 & 27 & 77 & 0.3 \\
Coalescence of multiple cracks (\ref{Sec:Percolation}) & 210000 & 0.3 & 2.7 & 4209 & 21047 & 0.006 \\
\bottomrule
\end{tabular}
\end{table}

\subsection{The surfing problem}\label{Sec:Surfing}

\begin{figure}[h]
	\centering
	\includegraphics[width=\textwidth]{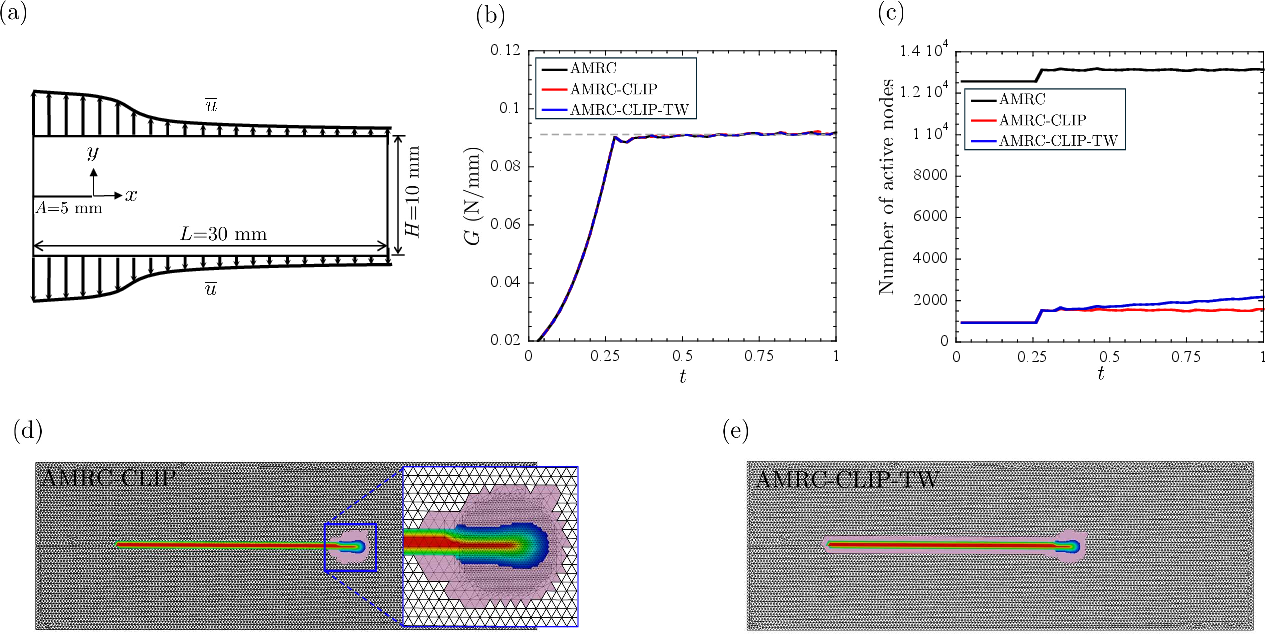}
	\caption{The surfing problem. (a) Geometry and boundary conditions. (b) Energy release rate computed from the $J$-integral as a function of time for AMRC, AMRC-CLIP, and AMRC-CLIP-TW; the dashed line is $G_c$. (c) Number of active nodes as a function of time. Phase field and mesh (d) for AMRC-CLIP, with the patch in pink and a close-up of the crack tip, and (e) for AMRC-CLIP-TW.}\label{Fig2}
\end{figure}

We begin with the surfing problem of Hossain et al.\ \cite{hossain2014}, a standard verification of Griffith-type propagation for phase-field models \cite{Tanne18, KBFLP20, KDK2025Comparison}. A strip of length $L=30$~mm and height $H=10$~mm contains an edge crack of length $A=5$~mm along its mid-plane (Fig.~\ref{Fig2}(a)). The vertical displacement of the top and bottom edges is that of the mode-I asymptotic solution \cite{KBFLP20, dahal2026amrc}. It is translated along the crack at 20~mm/s over $t\in[0,1]$~s in 50 steps, under plane stress, with at most 400 staggered iterations per step and $\texttt{tol}=10^{-7}$. The crack should then propagate steadily with loading, and the energy release rate $G$, computed from the $J$-integral along the boundary of the strip, should equal $G_c$.

The energy release rates computed with the AMRC-CLIP and AMRC-CLIP-TW, along with the reference solution with plain AMRC, are plotted in Fig.~\ref{Fig2}(b) and lie on top of each other. The average difference in $G$ between AMRC-CLIP and AMRC is 0.142\% and the AMRC-CLIP-TW result agrees exactly with AMRC. As discussed earlier, the small error in AMRC-CLIP is due to phase-field values not reaching their minimum in the wake; however, it is evident that the error is tiny.

The AMRC mesh has about $1.3\times10^4$ nodes, most of them in the base mesh. AMRC-CLIP solves the coupled problem on about $9\times10^2$ nodes before propagation and $1.6\times10^3$ nodes during propagation, an eightfold reduction, as shown in Fig.~\ref{Fig2}(c). This number remains fairly constant as the crack advances, because the patch travels with the tip. The close-up in Fig.~\ref{Fig2}(d) shows the active patch in pink that includes the refined window around the crack tip, and a small zone of coarse cells around it and in the wake. In AMRC-CLIP-TW, the pink patch encloses the entire crack band as shown in Fig.~\ref{Fig2}(e). Its number of active nodes grows linearly with the crack length, but slowly, to about $2.2\times10^3$ at $t=1$~s.

\begin{figure}[h]
	\centering
	\includegraphics[width=\textwidth]{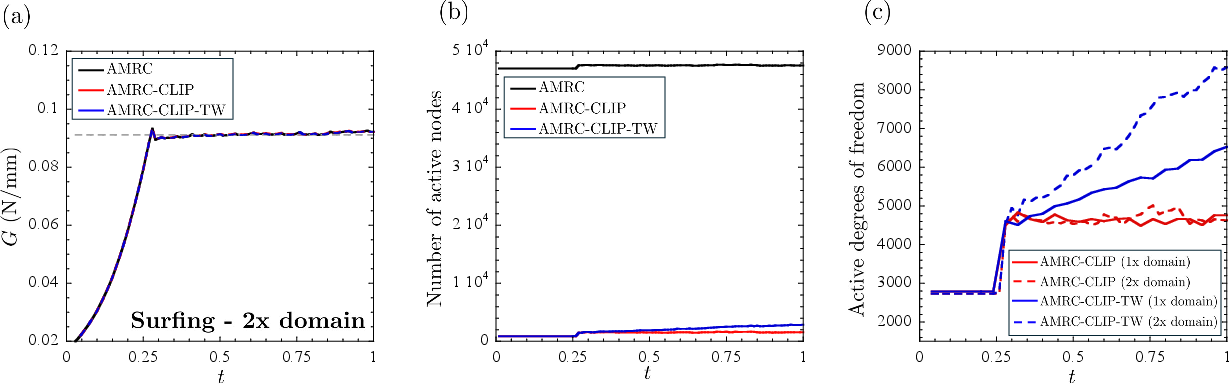}
	\caption{The surfing problem on a domain twice as large ($L=60$ mm, $H=20$ mm, $A=10$ mm). (a) Energy release rate and (b) number of active nodes as functions of the time for AMRC, AMRC-CLIP, and AMRC-CLIP-TW. (c) Number of active degrees of freedom of AMRC-CLIP (red) and AMRC-CLIP-TW (blue) on the original ($1\times$, solid) and the doubled ($2\times$, dashed) domain.}\label{Fig3}
\end{figure}

To examine how the reduction in active nodes scales with the size of the problem, we repeat the computation on a domain twice as large, whose base mesh has about four times as many nodes. $\varepsilon$, $h_0$, $\hmin$ are left unchanged. The energy release rates again coincide (Fig.~\ref{Fig3}(a)), with an average difference of 0.167\% for AMRC-CLIP. AMRC now operates on about $4.7\times10^4$ nodes (Fig.~\ref{Fig3}(b)), whereas the number of active nodes in AMRC-CLIP barely changes. Figure~\ref{Fig3}(c) compares the active degrees of freedom (for both displacement and phase fields) on the two domains. The AMRC-CLIP curves are very close, showing that, with this method, the number of active nodes depends only on the regularization length, not on domain size or crack length. In AMRC-CLIP-TW, the active degrees of freedom grow with the crack length, which grows to twice as long on the larger domain.

\begin{figure}[h]
	\centering
	\includegraphics[width=5.5in]{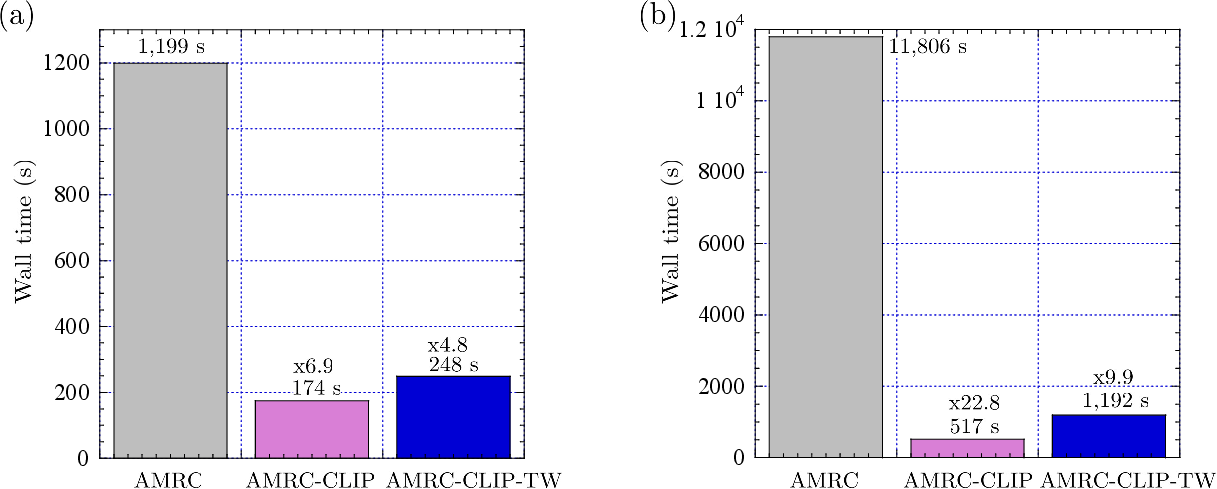}
	\caption{Wall time of AMRC, AMRC-CLIP, and AMRC-CLIP-TW for the surfing problem (a) on the original domain and (b) on the doubled domain; the speed-ups are relative to AMRC.}\label{Fig4}
\end{figure}

Fig.~\ref{Fig4} shows the savings in wall time with the proposed scheme. On the original domain, AMRC-CLIP takes 174~s, and AMRC-CLIP-TW takes 248~s, with speed-ups of 6.9 and 4.8 with respect to AMRC\footnote{The wall time of plain AMRC calculation can be significantly reduced by assuming a coarser base mesh size. Here for simplicity, we have assumed a base mesh size that is equal to the crack band mesh size.}. On the larger domain, the wall time of AMRC increases almost tenfold, to 11,806~s, whereas that of AMRC-CLIP only triples, to 517~s, a speed-up of 22.8. The reason the wall time triples even though number of active nodes remain constant is because of the work done at every move of the patch when $\mathbf{K}_{EE}$ is re-factorized: the work that scales with the size of the domain. 
AMRC-CLIP-TW takes 1,192~s, nearly 5 times larger than the wall time for the smaller domain. The advantage of AMRC-CLIP therefore grows with the size of the domain relative to the regularization length $\varepsilon$, which is precisely the regime of large structures in which phase-field computations are most expensive.

\subsection{Dynamic crack branching}\label{Sec:Branching}

\begin{figure}[h]
	\centering
	\includegraphics[width=\textwidth]{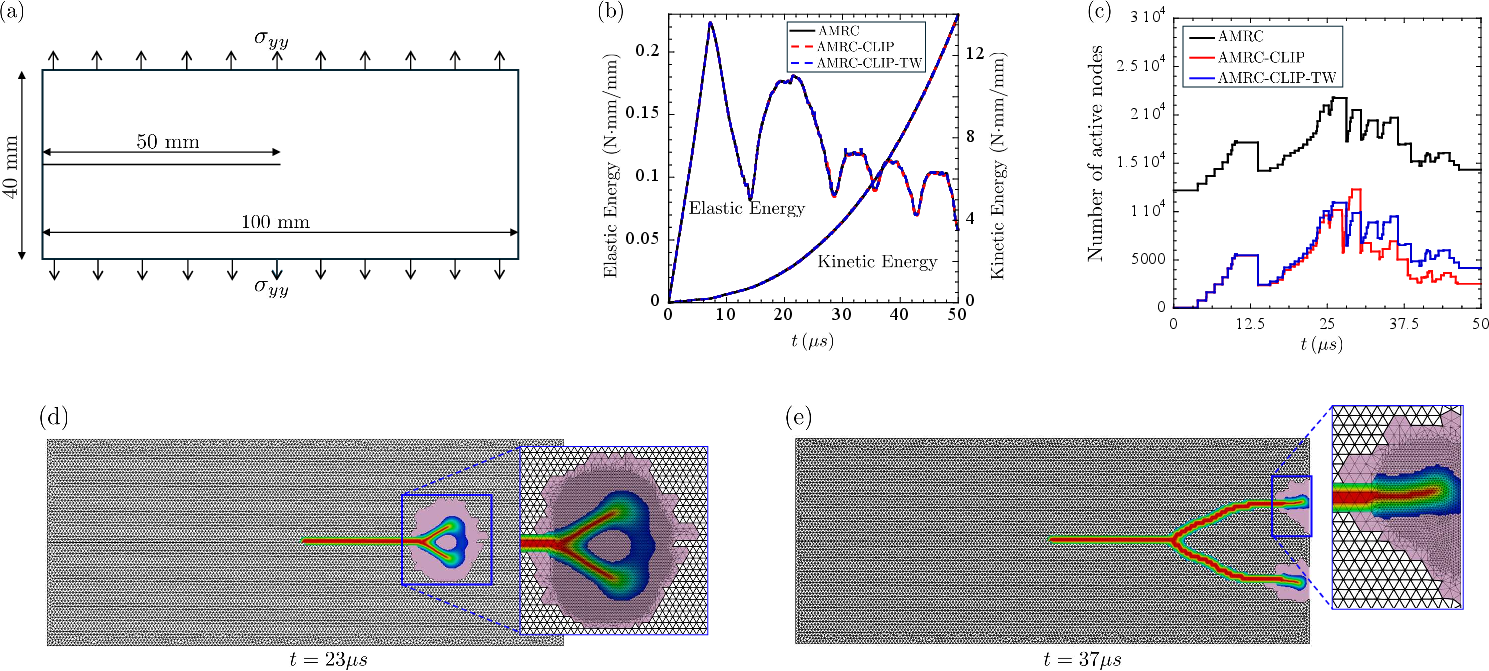}
	\caption{Dynamic crack branching. (a) Geometry and loading. (b) Elastic and kinetic energies as functions of time for AMRC, AMRC-CLIP, and AMRC-CLIP-TW. (c) Number of active nodes as a function of time. (d, e) Phase field and mesh of the AMRC-CLIP simulation at $t=23$~$\mu$s, during branching, and $t=37$~$\mu$s, after branching, with the patch in pink and close-ups of the crack tips.}\label{Fig5}
\end{figure}

The second problem tests AMRC-CLIP under dynamic loading to assess its robustness when cracks branch.
We consider the benchmark of Fig.~\ref{Fig5}(a) \cite{Song2008_ComparativeFE, borden2012phase, Bleyer2017_DynamicCrackVariational, dahal2026dynamic}, also studied with AMRC in \cite{dahal2026amrc}. A plate of 100~mm by 40~mm with an edge crack of length 50~mm along its mid-plane is loaded by a tensile traction $\sigma_{yy}=2$~MPa, applied on the top and bottom edges in 25~ns and then held constant. The governing equations (\ref{phase-field-dynamics}) are solved according to the finite element formulation discussed in Section \ref{Sec:FEM-Dyn}. The exterior is condensed with the effective stiffness (\ref{Eff-stiff-Dyn}).


The evolution of elastic and kinetic energies obtained with the three approaches are plotted in Fig.~\ref{Fig5}(b). 
They are almost indistinguishable throughout the initiation, propagation, and branching of the crack. The average differences between AMRC-CLIP and AMRC are 0.420\% in the elastic energy and 0.024\% in the kinetic energy. 

The number of active nodes is shown in Fig.~\ref{Fig5}(c). For AMRC-CLIP and AMRC-CLIP-TW, it remains at 0 until about 4~$\mu$s when crack starts to grow. 
The number of active nodes then grows to about $1.2\times10^4$ between 23 and 30~$\mu$s during branching, when the indicator flags the broad region in which the tip splits as shown in Fig.~\ref{Fig5}(d). 
At $t=37$~$\mu$s, the patch automatically splits into two disjoint components, each traveling with one branch as shown in Fig.~\ref{Fig5}(e), while the entire wake, including the branching point, lies in the exterior. The number of active nodes has decreased from the peak. No special treatment is needed to deal with branching.

At the end of the simulation, the patch has about $2.3\times10^3$ nodes for AMRC-CLIP and $4\times10^3$ for AMRC-CLIP-TW, whereas the AMRC mesh has between $1.2\times10^4$ and $2.2\times10^4$ nodes throughout. There is overall a speedup of 2.2x for the AMRC-CLIP compared to the reference AMRC. The wall time reduction is less compared to the surfing problem because the ratio of the number of nodes in the local patch to the exterior is smaller.

\subsection{Three-dimensional crack growth under in-plane shear}\label{Sec:Shear3D}

\begin{figure}[h]
	\centering
	\includegraphics[width=6in]{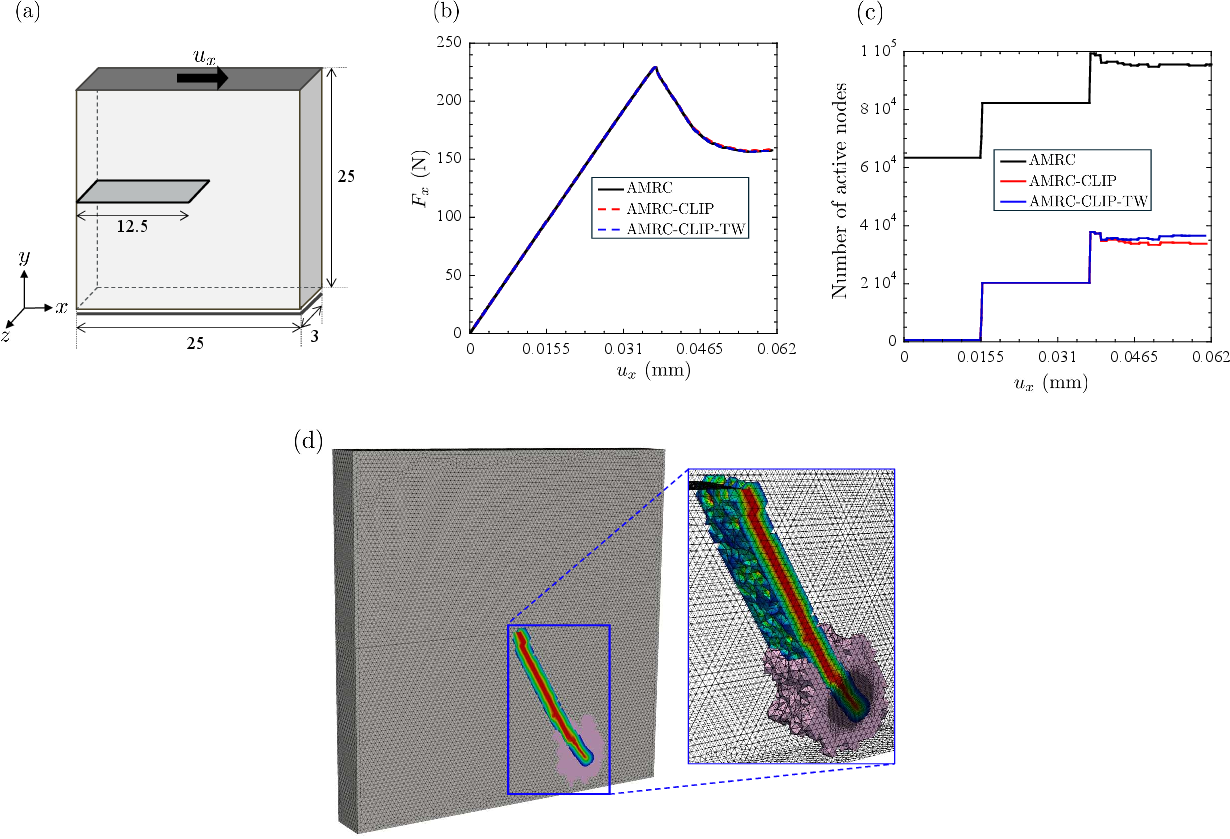}
	\caption{Three-dimensional crack growth under in-plane shear. (a) Geometry and loading, with dimensions in mm. (b) Reaction force on the top face as a function of the applied displacement for AMRC, AMRC-CLIP, and AMRC-CLIP-TW. (c) Number of active nodes as a function of the applied displacement. (d) Phase field and mesh of the AMRC-CLIP simulation, with the patch in pink and a close-up of the crack front.}\label{Fig6}
\end{figure}

The next problem we study is a three-dimensional version of the single-edge notched shear test \cite{miehe2010rate, KDK2025Comparison} to evaluate the robustness of the method in 3D. A plate of 25~mm by 25~mm and thickness 3~mm contains an edge crack of length 12.5~mm at mid-height, through its entire thickness (Fig.~\ref{Fig6}(a)). The bottom face is fixed, and the top face is displaced horizontally by $u_x$. The crack front is thus loaded in in-plane shear, and the crack kinks and propagates downward along an inclined path (Fig.~\ref{Fig6}(d)). In three dimensions, the patch must enclose a neighborhood of the entire crack front.

The results for the reaction force are shown in Fig.~\ref{Fig6}(b) for the three approaches.
In each case, the reaction force $F_x$ on the top face increases linearly to a peak of about 230~N at $u_x\approx0.037$~mm, when the crack starts to propagate, then decreases to about 157~N. The three curves coincide, and the average difference between AMRC-CLIP and AMRC is only 0.158\%. Figure~\ref{Fig6}(d) shows the patch around the front at the lower end of the crack, while the crack band in its wake lies in the exterior.

The number of active nodes is shown in Fig.~\ref{Fig6}(c). Until $u_x\approx0.0155$~mm, AMRC operates on the base mesh of about $6.3\times10^4$ nodes, while the patch of AMRC-CLIP and AMRC-CLIP-TW is empty. As the phase field develops at the crack tip, the indicator flags the front, and AMRC refines a window along it. Active nodes in the patch grow to $2.0\times10^4$ nodes.
There is another jump in the patch nodes when a phase field crack fully nucleates from the pre-existing crack at $u_x\approx0.036$~mm.
Once the crack propagates, the AMRC mesh settles at about $9.5\times10^4$ nodes while that of AMRC-CLIP settles at about $3.4\times10^4$ nodes
The overall reduction in wall time for AMRC-CLIP is again around 2x.

\subsection{Coalescence of multiple cracks}\label{Sec:Percolation}

\begin{figure}[h]
	\centering
	\includegraphics[width=\textwidth]{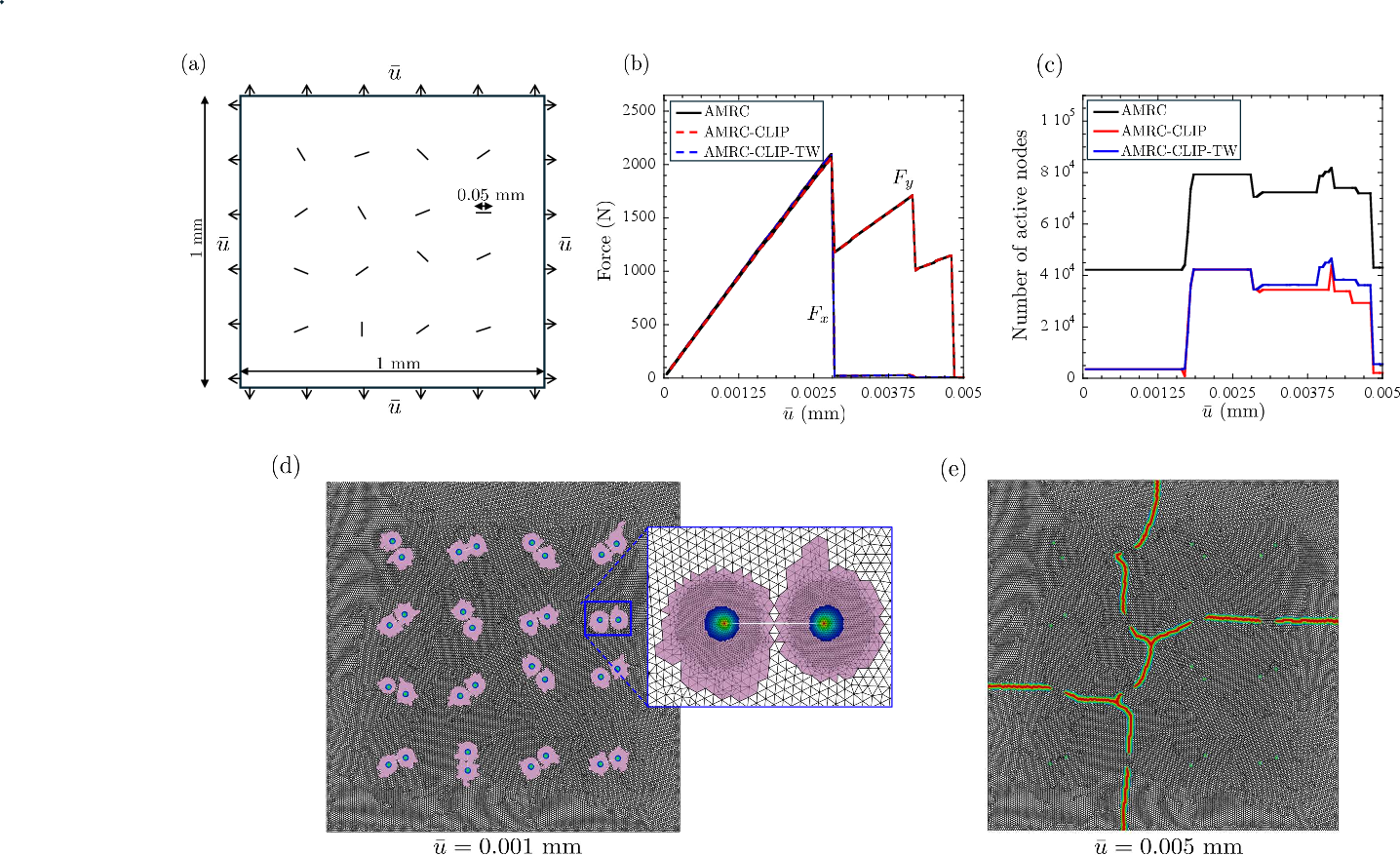}
	\caption{Growth and coalescence of multiple cracks in a unit cell, adapted from \cite{geelen2020extended}. (a) Geometry and loading. (b) Reactions $F_x$ on the right edge and $F_y$ on the top edge as functions of the edge displacement $\bar{u}$ for AMRC, AMRC-CLIP, and AMRC-CLIP-TW. (c) Number of active nodes as a function of $\bar{u}$. (d) Phase field and mesh of the AMRC-CLIP simulation at $\bar{u}=0.001$~mm, with the patch in pink and a close-up of one crack. (e) Final crack pattern obtained from AMRC-CLIP.}\label{Fig7}
\end{figure}

The last problem tests AMRC-CLIP on a unit-cell problem, adapted from Geelen et al.~\cite{geelen2020extended}, where many crack tips are active simultaneously, and the cracks coalesce. 
We consider a unit cell of 1~mm by 1~mm under plane strain conditions with 16 randomly distributed microcracks of length 0.05~mm (Fig.~\ref{Fig7}(a)). Each crack is a geometrical slit in the mesh, with $v=0$ assigned at its tips. All four edges are displaced outward by $\bar{u}$, increased quasi-statically to 0.005~mm, with the tangential displacement free. As the cracks reach the loaded edges, the patch can contain nodes with prescribed displacements, which are imposed as Dirichlet conditions on the patch problem.

The reactions on the right and top edges, $F_x$ and $F_y$ respectively, are shown in Fig.~\ref{Fig7}(b). Both increase linearly up to $2.1\times10^3$~N/mm at $\bar{u}=0.0028$~mm. The cracks of the second column then coalesce into a single crack that runs from the bottom to the top edge. As a result, $F_x$ drops to zero, and $F_y$ to $1.2\times10^3$~N. As the loading continues, horizontal cracks connect this macro crack to the left edge, at $\bar{u}=0.0042$~mm, and to the right edge, at $\bar{u}=0.00485$~mm, where the domain fails completely, and $F_y$ also drops to zero. The final crack path is shown in Fig.~\ref{Fig7}(e). The average differences between AMRC-CLIP and AMRC are 0.0017\% in $F_x$ and 0.024\% in $F_y$. Table~\ref{tab:errors} summarizes the errors for all four problems.

The number of active nodes is shown in Fig.~\ref{Fig7}(c). The initial patch of AMRC-CLIP and AMRC-CLIP-TW is built around all 32 tips and has $3.5\times10^3$ nodes, whereas AMRC operates on a mesh of $4.2\times10^4$ nodes. As the cracks grow and coalesce, the AMRC mesh grows to $7\times10^4$ to $8\times10^4$ nodes; however, the AMRC-CLIP patch has about half as many nodes, between $2.9\times10^4$ and $4.4\times10^4$. At the end of the simulation, the AMRC-CLIP patch shrinks to $2\times10^3$ nodes, and the AMRC-CLIP-TW patch to $6\times10^3$ nodes.

\begin{table}[h]
\centering
\caption{Average relative difference of the global responses computed from AMRC-CLIP with respect to those computed from AMRC. AMRC-CLIP-TW is exact with respect to AMRC.}
\label{tab:errors}
\small
\begin{tabular}{llc}
\toprule
Problem & Response & Difference \\
\midrule
Surfing, original domain (\ref{Sec:Surfing}) & Energy release rate $G$ & 0.142\% \\

Surfing, doubled domain (\ref{Sec:Surfing}) & Energy release rate $G$ & 0.167\% \\
Dynamic branching (\ref{Sec:Branching}) & Elastic energy & 0.420\% \\
 & Kinetic energy & 0.024\% \\
3D in-plane shear (\ref{Sec:Shear3D}) & Reaction force $F_x$ & 0.158\% \\
Coalescence of multiple cracks (\ref{Sec:Percolation}) & Reaction force $F_x$ & 0.0017\% \\
 & Reaction force $F_y$ & 0.024\% \\
\bottomrule
\end{tabular}
\end{table}

\section{Final comments}\label{Sec:Conclusions}

This work has introduced AMRC-CLIP, an extension of the adaptive mesh refinement and coarsening framework that localizes not only the spatial discretization but also the computational effort associated with the repeated solution effort in phase-field fracture simulations. The central observation is that, once the phase field and mesh in the inactive exterior are fixed, the linear response of this region can be condensed onto the boundary of a small active patch surrounding the crack tip. The corresponding factorization and Schur complement are computed only when the patch moves and can otherwise be reused throughout the staggered iterations. 

The principal advantage of the method is expected for large structures whose dimensions are much larger than the intrinsic fracture length scales. In AMRC-CLIP, the size of the repeatedly solved problem is controlled primarily by the size of the active crack-tip region rather than by the size of the overall structure or the accumulated crack length. The expensive global factorization is therefore amortized over the many hundreds of staggered iterations performed while the crack remains within the same patch. This behavior is clearly evident from the results for the surfing problem. For that problem, doubling the dimensions of the domain increases the speedup over AMRC from $6.9$ to $22.8$, while the number of active degrees of freedom of AMRC-CLIP remains essentially unchanged. The advantage should therefore increase further as the ratio between structural dimensions and the regularization length increases. This localization of the nonlinear fracture computation may also be particularly useful in applications requiring many repeated fracture calculations, including topology optimization, computational homogenization, and multiscale modeling. More broadly, because the active problem is posed on a local region whose dimensions are governed primarily by intrinsic material length scales rather than the global geometry, the framework provides a natural route toward constructing generalizable surrogate models for the local fracture problem, while retaining a conventional FE solver for the surrounding structure \cite{Karniadakis2022interfacing}.

There is scope for improving the efficiency of the patch construction itself. In the present work, the patch size is prescribed relative to $\varepsilon$. An adaptive choice could instead balance two competing costs. A larger patch increases the cost of every staggered iteration but allows the crack to propagate farther before the exterior must be refactorized. On the other hand, a smaller patch reduces the local problem size but requires more frequent patch moves and factorizations. The optimal patch size should therefore depend, among other factors, on the rate of crack propagation, the number of staggered iterations required per increment, and the relative cost of exterior factorization and local solution. Dynamically adjusting the patch size, possibly with a trained neural network, according to these quantities provides a natural next step for the method.

For geometrically complex structures and to parallelize the computations better, an alternative direction is to combine the present moving active-region concept with global--local enriched finite element formulations \cite{Duarte2007GlobalLocal, Duarte2008GlobalLocal, Duarte2010ThreeDimensional}. For example, the X/G-FEM global--local framework of Geelen et al.~\cite{geelen2020extended} enriches a global finite element description using the solution of a local phase-field problem. In their approach, the local fracture domain typically expands as the crack grows. Our approach suggests instead that the high-resolution local domain could remain of approximately fixed size and translate with the active crack tip. The crack wake could then be represented in the global model through a discontinuous or Heaviside-type enrichment. Such a representation may be particularly attractive when explicit access to the crack faces is desirable, for example to describe contact, fluid interaction, or penetration by an indenter. 

Non-intrusive global--local approaches provide another promising alternative. In the formulations of Gerasimov et al.~\cite{gerasimov2018globallocal} and Noii et al.~\cite{noii2020adaptive}, the global problem remains largely unchanged while the nonlinear fracture response is resolved only in a local domain and communicated to the global model through interface quantities. Such approaches are attractive because they can retain the existing global finite element operator. 
They require, however, iterative exchange between the global and local problems. The relative efficiency of different global--local formulations and the direct Schur-complement condensation for large-scale phase field fracture problems remains to be investigated.

\section*{Acknowledgements}

\noindent The authors would like to acknowledge the financial support from the National Science Foundation, United States, through the grants CMMI-2404808 and 2541441. The authors benefited from discussions with Abhinav Gupta (Vanderbilt University).

\section*{Data Availability}
The entire source code is available at \url{https://github.com/Aditya-Kumar-Lab-GT/AMRC-CLIP}.

\bibliographystyle{unsrt}
\bibliography{ref}

\end{document}

%% file: definitions.tex
\def\bfu{{\bf u}}

\def\bfE{{\bf E}}

\def\bfI{{\bf I}}

\def\bfS{{\bf S}}

\def\bfX{{\bf X}}
\def\bfF{{\bf F}}

\def\eps{\varepsilon}

\def\e0{\varepsilon_0}

\def\s0{\sigma_0}
\def\v{\varphi}

\def\sts{\sigma_{\texttt{ts}}}
\def\scs{\sigma_{\texttt{cs}}}
\def\shs{\sigma_{\texttt{hs}}}
\def\sss{\sigma_{\texttt{ss}}}
\def\de{\delta^\varepsilon}
\def\ce{c_{\texttt{e}}}

\DeclareMathAlphabet{\mathsfit}{T1}{\sfdefault}{\mddefault}{\sldefault}
\SetMathAlphabet{\mathsfit}{bold}{T1}{\sfdefault}{\bfdefault}{\sldefault}